# Soil cracking modelling using the mesh-free SPH method

H.H. Bui[1], G.D. Nguyen[2], J. Kodikara[1] and M. Sanchez[3]

[1,3]Department of Civil Engineering, Monash University, Clayton, Victoria 3800; PH (+61) 9905-2599; FAX (+61) 9905-4944; email: ha.bui@monash.edu
[2]School of Civil, Environmental & Mining Engineering, The University of Adelaide, Adelaide, South Australia, 5005; PH (+61) 8-8313-2259; FAX (+61) 8-8313-4359; email: g.nguyen@adelaide.edu.au
[3]Department of Civil Engineering, Texas A&M University, Texas, 3136; PH (+1) 979-862-6604; FAX (+1) 997-862-7969; email: msanchez@civil.tamu.edu

**ABSTRACT**

The presence of desiccation cracks in soils can significantly alter their mechanical and hydrological properties. In many circumstances, desiccation cracking in soils can cause significant damage to earthen or soil supported structures. For example, desiccation cracks can act as the preference path way for water flow, which can facilitate seepage flow causing internal erosion inside earth structures. Desiccation cracks can also trigger slope failures and landslides. Therefore, developing a computational procedure to predict desiccation cracking behaviour in soils is vital for dealing with key issues relevant to a range of applications in geotechnical and geo-environment engineering. In this paper, the smoothed particle hydrodynamics (SPH) method will be extended for the first time to simulate shrinkage-induced soil cracking. The main objective of this work is to examine the performance of the proposed numerical approach in simulating the strong discontinuity in material behaviour and to learn about the crack formation in soils, looking at the effects of soil thickness on the cracking patterns. Results show that the SPH is a promising numerical approach for simulating crack formation in soils

## 1   INTRODUCTION

Desiccation cracks are formed by the shrinkage of soils due to moisture loss. The presence of cracks can induce significant changes in the mechanical, hydrological, physico-chemical and thermal properties of soils (Kodikara and Costa 2013; Konrad and Ayad 1997; Miller et al. 1998; Peron et al. 2007; Peron et al. 2009; Peron et al. 2009). This can lead, for example, to damage of lightly loaded structures (e.g., residential houses) or shallow-buried structures (e.g., gas and water pipelines), progressive slope/dam failures, cracking in road pavements, and the leakage of deep nuclear waste/hazardous gasses from soils (Kodikara and Costa 2013). Therefore, understanding the mechanism of desiccation cracking behaviour in clayey soils is vital for dealing with key issues relevant to a range of applications in numerous disciplines such as geotechnical engineering, geo-environment engineering, transport engineering, mining and resource engineering, agricultural engineering and soil science. Over the last few decades, there has been a substantial research effort around the world directed at studying and modelling desiccation-induced shrinkage cracks in soils. Early experimental work on modelling of desiccation cracking of soils mostly used rectangular boxes (Miller et al. 1998; Yesiller et al. 2000). While these experiments provide valuable data, the results were far too complex for detailed analysis or numerical modelling (Kodikara and Costa 2013). Long moulds tests have been subsequently introduced to investigate desiccation cracking of soils where crack patterns are controlled so that they develop parallel to the longitudinal direction (Costa et al. 2008; Peron et al. 2009). Particle image velocimetry (PIV) has been also adopted to capture the complete picture of crack evolution as well as propagation (Costa et al. 2008). These advancements have improved our understanding of the fracture behaviour of clayey soils. However, mechanisms and variables associated with drying and especially with desiccation shrinkage and cracking are still far from fully understood.

The numerical modelling of desiccation cracking in clayey soils has progressed during the last few decades. Research in this area can be classified into two main categories, that is, continuum and discontinuum numerical approaches. The continuum approach is based on the finite element method (FEM) with continuum constitutive models that relate stress and strain through the application of classical plasticity theory (Hallett and Newson 2005; Hu et al. 2008; Peron et al. 2007; Peron et al.

2009; Vogel et al. 2005; Yoshida and Adachi 2004); while the discontinuum approach is based on the discrete element method (DEM) which tracks the motion of a large number of grains (either spherical or irregular shaped) using inter-particle contact laws that relate forces/torques to relative movement between two contacting grains (Amarasiri and Kodikara 2011; Amarasiri and Kodikara 2013; Amarasiri et al. 2011; Amarisiri et al. 2014; Kodikara et al. 2004; Peron et al. 2009; Sima et al. 2013). The major disadvantages of the continuum approach arise from the use of the FEM, which is usually difficult to model crack initiation and propagation. On the other hand, the discontinuum approach seems to be a promising approach to model desiccation cracking in soils; but, it is unable to predict multi-physical processes in soil desiccation cracking (e.g., moisture evaporation, heat and mass exchanges). Another powerful continuum-based approach is the mesh-free smoothed particle hydrodynamics (SPH) method (Gingold and Monaghan 1977; Monaghan 2012), in which continuum equations are solved at material points. This method allows the use of any constitutive model and/or failure criterion for material points representing fluid, grains, bonds, and thus provides an excellent means to model desiccation-induced soil cracking processes. The SPH method has been successfully applied to various engineering applications, such as: fluid dynamics (Monaghan 1994); multi-phase flows (Hu and Adams 2006; Monaghan and Rafiee 2013); flow through porous media (Bui and Fukagawa 2011; Holmes et al. 2011); heat conduction (Cleary and Monaghan 1999); and failure of geomaterials (Bui and Fukagawa 2013; Bui et al. 2008; Bui et al. 2009; Bui et al. 2011; Bui and Khoa 2011; Bui et al. 2014; Bui et al. 2007; Bui et al. 2008; Yaidel et al. 2014). However, a SPH model with the ability to simulate desiccation cracks in soil has not been developed. In this paper, the potential application of the SPH method to simulate desiccation cracks in soil will be investigated.

## 2 SIMULATION APPROACHES

### 2.1 Soil deformation in SPH

The basic equations used to describe the motion of soil in the SPH framework are the continuity equation and the momentum equation (Bui et al. 2008). The continuity equation describes the change in density and void ratio of soil undergoing large deformation, while the momentum equation simulates soil deformation subjected to external loading. These two equations are written as follows,

$$\frac{d\rho}{dt} = -\nabla \cdot \mathbf{v} \tag{1}$$

$$\rho \frac{d\mathbf{v}}{dt} = \nabla \cdot \boldsymbol{\sigma} + \rho \mathbf{g} + \mathbf{f}_{ext} \tag{2}$$

where $\mathbf{v}$ is the velocity vector of the soil particles; $\rho$ is the density; $\boldsymbol{\sigma}$ is the total stress tensor, taken negative for compression; $\mathbf{g}$ is the acceleration due to gravity; and $\mathbf{f}_{ext}$ represents the additional external forces.

The total stress tensor of soil ($\boldsymbol{\sigma}$) is normally composed of the effective stress ($\boldsymbol{\sigma}'$) and the pore-water pressure (or matrix suction), following Terzaghi's concept of effective stress. Because the effects of water pressure are not considered in this paper, the total stress tensor and the effective stress are identical throughout this paper and can be computed using any material constitutive model. It is noted that if the soil density is assumed constant throughout the numerical analysis, which is the case for most FEM analyses, the continuity equation could be omitted. For large deformation analyses, however, it is recommended to take into consideration of the density change of soil during the failure process. This can be done by solving the continuity equation and updating the density during the numerical analysis.

Within the SPH framework, the partial differential form of equations (1) and (2) can be discretised in the following way (Bui et al. 2008),

$$\frac{d\rho_a}{dt} = \sum_{b=1}^{N} m_b \left(v_a^\alpha - v_b^\alpha\right) \cdot \frac{\partial W_{ab}}{\partial x_a^\alpha} \tag{3}$$

$$\frac{dv_a^\alpha}{dt} = \sum_{b=1}^{N} m_b \left(\frac{\sigma_a^{\alpha\beta}}{\rho_a^2} + \frac{\sigma_b^{\alpha\beta}}{\rho_b^2} + C_{ab}^{\alpha\beta}\right) \frac{\partial W_{ab}}{\partial x_a^\beta} \nabla + g_a^\alpha + f_{ext \to a}^\alpha \tag{4}$$

where $\alpha$ and $\beta$ denote Cartesian components *x*, *y*, *z* with the Einstein convention applied to repeated indices; *a* indicates the particle under consideration; $\rho_a$ and $\rho_b$ are the densities of particle *a* and *b* respectively; *N* is the number of "neighbouring particles", *i.e.*, those in the support domain of particle *a*; $m_b$ is the mass of particle *b*; *W* is the kernel function, which is chosen to be the cubic-spline function (Monaghan and Lattanzio 1985); *C* is the stabilization term, which consists of an artificial viscosity (Monaghan 2005) and artificial stress (Gray et al. 2001), employed to remove the numerical and tensile instability issues associated with the SPH method (Bui and Fukagawa 2013); and $f_{ext \to a}$ is the external force acting on particle *a*. It is noted that the full extension of the artificial stress approach (Gray et al. 2001) to three-dimensions has not been undertaken in the literature. In this paper, for the first time, the artificial stress is fully extended to 3D and successfully applied to remove the 3D tensile instability problem in SPH simulations. Details of the extension can be found in (Yaidel et al. 2014). Finally, in order to complete the above system of governing equations, a mechanical constitutive model needs to be specified to calculate the stress tensor appeared in equation (4). This will be discussed in the following section.

## 2.2   Tension damage constitutive model

A mechanical constitutive model based on fracture and damage theory can be used to describe the degradation process of geomaterials subjected to loading. In this paper, in order to demonstrate the application of the SPH method to simulate soil cracking, a simple tension damage model was adopted. The model tracks the mechanical degradation of material via a scalar damage variable (*d*), which, in general, varies in the range between "0" for intact material and "1" for completed damage material. The degradation of materials in these damage models is associated with the appearance of micro-fissures when the loading increases above a given threshold. The tension damage model is described using the following constitutive model,

$$\boldsymbol{\sigma} = (1-d) \cdot \bar{\boldsymbol{\sigma}} \tag{5}$$

where $\boldsymbol{\sigma}$ is the damaged stress tensor, $\bar{\boldsymbol{\sigma}}$ is the elastic stress (i.e. elastic stress of the intact soil), and *d* is a scalar damage which is zero for intact and 1 for completely damaged materials. The discretization form of the rate of change of equation (5) can be rewritten as follows,

$$\frac{d\sigma_a^{\alpha\beta}}{dt} = (1-d_a)(2G_a \dot{e}_a^{\alpha\beta} + K_a \dot{\varepsilon}_a^{\gamma\gamma} \delta_a^{\alpha\beta}) \tag{6}$$

with $\dot{\varepsilon}_a^{\alpha\beta}$ being the strain-rate tensor, $\dot{e}_a^{\alpha\beta}$ the deviatoric strain-rate tensor, *G* the elastic shear modulus, *K* the elastic bulk modulus, and $\delta_a^{\alpha\beta}$ the Kronnecker delta function. The strain-rate tensor is computed from,

$$\dot{\varepsilon}_a^{\alpha\beta} = \frac{1}{2}\left[\sum_{b=1}^{N}\frac{m_b}{\rho_b}(v_b^\alpha - v_a^\alpha)\frac{\partial W_{ab}}{\partial x_a^\beta} + \sum_{b=1}^{N}\frac{m_b}{\rho_b}(v_b^\beta - v_a^\beta)\frac{\partial W_{ab}}{\partial x_a^\alpha}\right] \tag{7}$$

When considering a large deformation problem, a stress rate that is invariant with respect to rigid-body rotation must be employed for the constitutive relations. In the current study, the Jaumann stress rate, $\hat{\dot{\sigma}}_a^{\alpha\beta}$, is adopted:

$$\hat{\dot{\sigma}}_a^{\alpha\beta} = \dot{\sigma}_a^{\alpha\beta} - \sigma_a^{\alpha\gamma}\dot{\omega}_a^{\beta\gamma} - \sigma_a^{\gamma\beta}\dot{\omega}_a^{\alpha\gamma} \tag{8}$$

where $\dot{\omega}_a^{\alpha\beta}$ is spin-rate tensor computed by

$$\dot{\omega}_a^{\alpha\beta} = \frac{1}{2}\left[\sum_{b=1}^{N}\frac{m_b}{\rho_b}(v_b^\alpha - v_a^\alpha)\frac{\partial W_{ab}}{\partial x_a^\beta} - \sum_{b=1}^{N}\frac{m_b}{\rho_b}(v_b^\beta - v_a^\beta)\frac{\partial W_{ab}}{\partial x_a^\alpha}\right] \tag{9}$$

As a result, the stress-strain relationship for the current soil model becomes

$$\frac{d\sigma_a^{\alpha\beta}}{dt} = (1-d_a)(\sigma_a^{\alpha\gamma}\dot{\omega}_a^{\beta\gamma} + \sigma_a^{\gamma\beta}\dot{\omega}_a^{\alpha\gamma} + 2G_a\dot{e}_a^{\alpha\beta} + K_a\dot{\varepsilon}_a^{\gamma\gamma}\delta_a^{\alpha\beta}) \quad (10)$$

Finally, in order to solve equation (10), a damage criterion together with damage evolution law need to be adopted to describe the degradation process of geomaterials subjected loading. For a rigorous numerical modelling approach, damage-plasticity models and/or cohesive crack models should be applied to simulate crack initial and propagation, which requires further development. For the sake of simplicity, this paper has adopted an assumption that the soil is completely damaged after their maximum principal stresses reach the tensile damage stresses. Accordingly, the damage criterion which defines the elastic domain can be written as follows,

$$f_D = \sigma_I - \sigma_t \leq 0 \quad (11)$$

where $\sigma_t$ is the tensile damage stress and $\sigma_I$ is the principal stress in *I* direction. Accordingly, the procedure to compute damage stresses is as follows,

1) The 3D Cartesian stress tensor of each soil particle will be first translated to the principal stresses by computing the eigenvalues and eigenvectors of the stress tensor.

2) On the principal stress plane, the principal stresses are checked against the maximum tensile damage stress, i.e. equation (11). If the principal stresses exceed the tensile damage stress, the damage scalar is set to d = 1, and the principal stress is set to zero (completely damaged).

The modified principal stresses are finally translated back to the Cartesian stresses via the calculation of eigenvalues and eigenvectors of the stress tensor. These Cartesian stresses are subsequently used to compute soil motion in equation (2).

## 3   NUMERICAL APPLICATIONS

We present the application of the proposed numerical method to model soil cracking in long rectangular moulds, looking at the cracking patterns and the effect of soil thickness. Our primary intention is not to reproduce exactly the experimental results, but to verify whether or not the proposed technique is able to capture the essential mechanism of soil cracking behaviour. The soil material adopted in this paper is mining waste, which was previously reported by Rodiguez et al. (Rodriguez et al. 2007). The tensile strength and elastic properties of this material have been recently investigated by Sanchez et al. (Marcelo Sánchez 2014) and are summarised in Table 1.

*Table 1:* Material properties

| Element | E | v | ρ | $\sigma_t$ |
|---|---|---|---|---|
| Soil | 4MPa | 0.2 | 1330kg/m$^3$ | 4kPa |

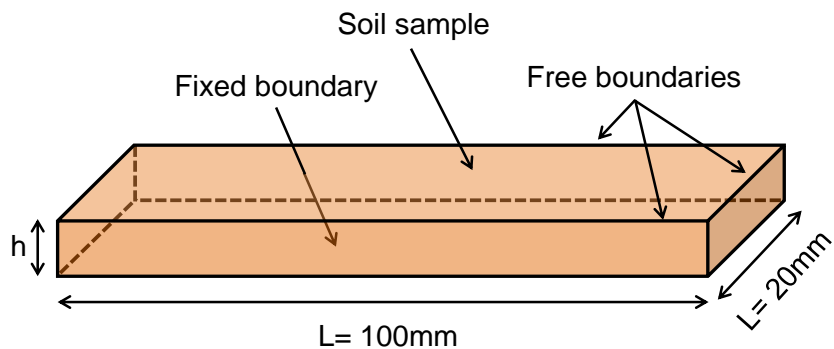

*Figure 1.* Initial geometry and boundary conditions of the numerical model

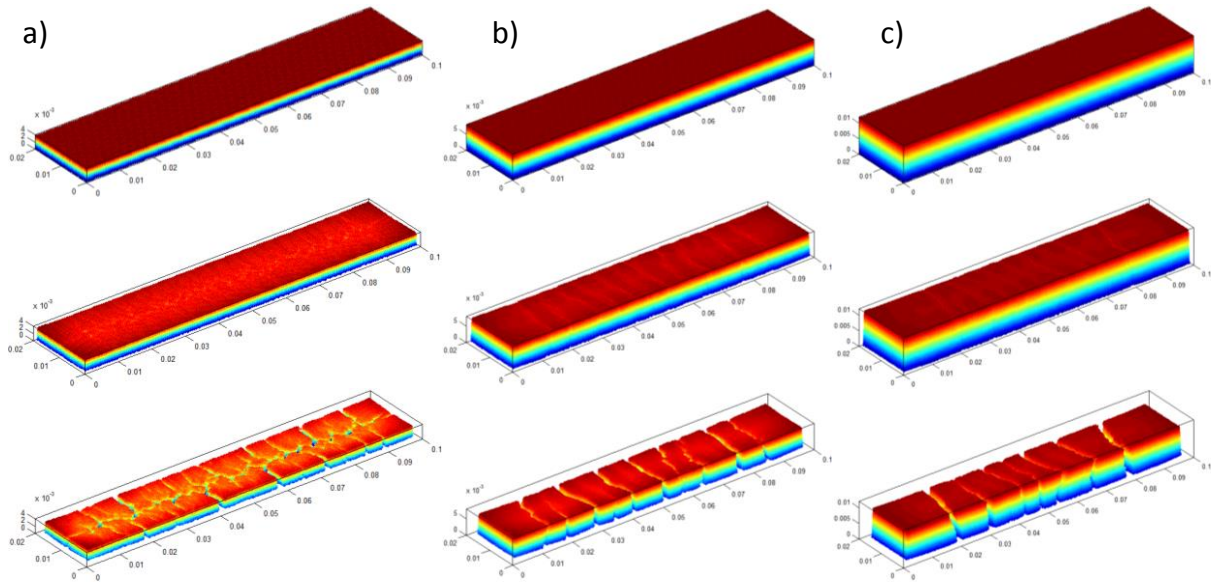

*Figure 2.* Evolution of the cracks during the analysis for different soil thickness: a) h = 4mm, b) h = 8mm and c) h 12mm. The bottom images show the final predicted cracks.

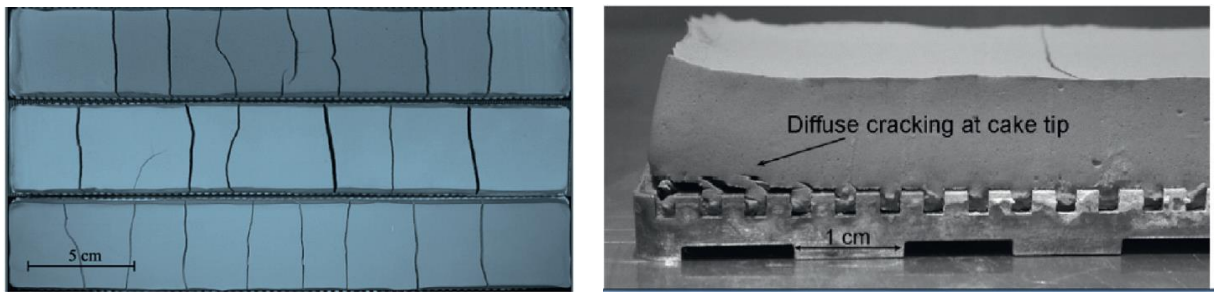

*Figure 3.* Experimental results conducted by (Peron et al. 2009).

Figure 1 shows the geometry and boundary conditions for the 3D numerical model. The model has dimension of 0.1m in length, 0.02m in width and h in height. The value of h is subjected to change from h = 0.004m to h = 0.008m and h = 0.012m in order to investigate the effect of soil thickness on the soil cracking patterns. Boundary conditions have been fully restrained in all direction at the bottom boundary and fully-free in all other boundaries. In the SPH method, the fully restrained boundary condition was modelled using virtual particles with a no-slip boundary condition (Bui et al. 2008). The total number of particles used for each simulation is 64,000 particles, 128,000 particles and 192,000 particles for h = 0.004m to h = 0.008m and h = 0.012m, respectively.

The next step is to simulate moisture evaporation induced soil shrinkage. A large number of desiccation tests on the current soil sample were carried out by (Rodriguez et al. 2007) using circular and rectangular plates. The tests were conducted on soil samples with initial moisture contents of 40%-50% in a controlled environmental chamber of a fairly constant temperature of $21^oC$ and a relative humidity close to 65%. In most cases, test results showed a maximum contraction deformation of approximately 5%. Accordingly, the same deformation rate will be adopted in our simulation without solving the moisture flow problem. The contraction rate was introduced to the numerical simulation by gradually applying shrinkage stresses to all soil particles. The shrinkage stresses were calculated as the product of the increment of the volumetric strain tensor and the elastic Young's modulus.

Figure 2 shows the desiccation-induced soil cracking process of three soil samples with different layer thickness. Soils undergo elastic shrinkage during the first stage without forming cracks or tensile instability (top images). It is worth mentioning that tensile instability is a well-known issue associated with the SPH method when simulating materials subjected to tensile behaviour. In most cases, SPH particles tend to attract each other under tensile stress, which results to form an unphysical clumping in the SPH particles (Gray et al. 2001). The problem could be completely removed in most 2D simulations using the artificial stress approach (Bui et al. 2008). However, the extension of this approach to 3D condition has not been undertaken in the past. In order to test the effectiveness of the 3D artificial stress algorithm adopted in this paper, we conducted several numerical tests without adopting the tensile instability treatment. In such tests, soil particles clumped and physically unrealistic cracks appeared on the soil surface before reaching their damage stages. This result suggests that the artificial stress method works well for the 3D problem.

As the shrinkage loading stress increases, the redistribution of tensile stresses takes place. The stresses concentrate on the surface due to the constraint imposed at the bottom boundary and keep increasing until the principal stresses reach the tensile soil strength, at which cracks start to develop and again stress redistribution takes place (middle images). The bottom images in Figure 2 show the final cracking pattern obtained from the simulation for 5% contraction deformation. It can be seen that the numerical simulations could reproduce satisfactory results similar to those observed in the experiments conducted by (Peron et al. 2009) as shown in the left image of Figure 3. Comparing the numerical results of the three soil samples, it can be seen that the number of fragmentation cracks reduces as the soil thickness increase. The spacing between cracks is also controlled by the soil thickness. These simulation results are consistent with the experimental finding reported by (Rodriguez et al. 2007) and with the numerical simulations conducted by (Marcelo Sánchez 2014) using a mesh fragmentation technique. However, compared to the mesh fragmentation technique (Marcelo Sánchez 2014), the SPH method is more robust as it requires no interface element to handle cracks, which significantly reduces computational costs.

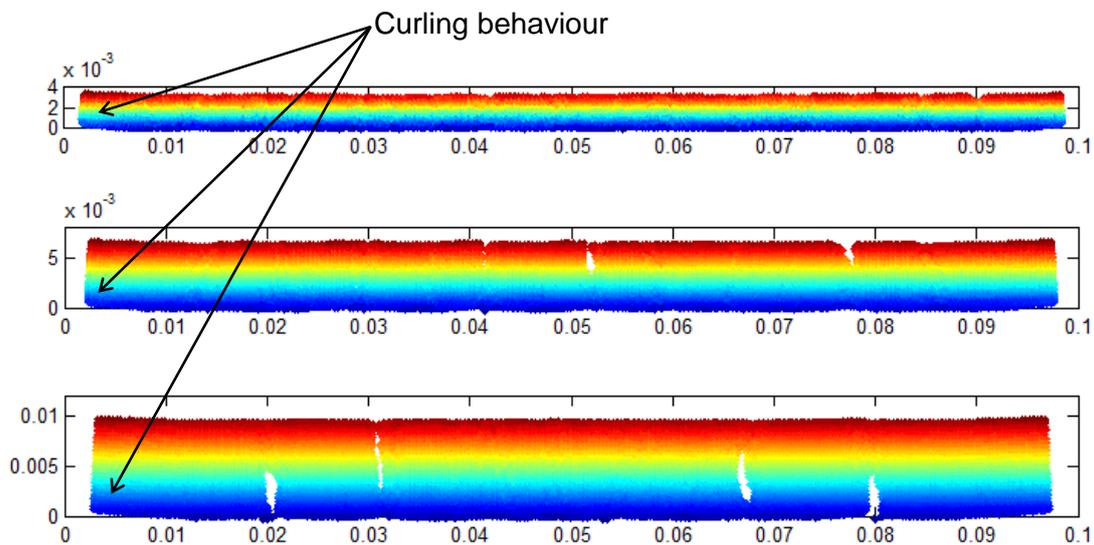

*Figure 4.* Curling deformation in desiccation cracking

Finally, we examine the capacity of SPH to model curling behaviour, which was reported by (Kodikara et al. 2004) and (Peron et al. 2009) through experimental investigations. As can be seen from Figure 4, the curling behaviour of the soil samples at the left and right tips of the soil sample could be simulated well using the proposed numerical technique. The obtained numerical crack patterns reflect well the experimental observations by (Peron et al. 2009) as shown in the right image of Figure 3.

Despite the use of a very simple brittle damage model, the proposed SPH models could qualitatively reproduce the experimental observations in all numerical simulations. We note that the underlying physics of soil desiccation is missing in this simple damage model and this is one of the further developments in our future work, besides the developments of SPH algorithms. In addition, continuum damage models describing fracture is subjected to several pathological issues associated with the loss of stability of the governing differential equations. Discretisation-dependent numerical results are a direct consequence of this stability issue. In this case, further development of a new continuum model with embedded cohesive crack and able to produce discretisation-dependent results (Nguyen et al. 2014) is a logical next step for the current SPH-based modelling and simulation.

## 4   CONCLUSION

This paper has presented the application of the SPH method to simulate crack formation in soils. A simple tension damage model was proposed and successfully implemented in the SPH code to simulate soil behaviour. In order to remove the tensile instability problem in SPH, the artificial stress approach was successfully extended to the 3D conditions. The proposed method was then applied to simulate shrinkage deformation induced soil cracking and to investigate the effects of the soil thickness on cracking pattern. Early numerical results showed good agreement with the experimental observation reported in the literature, that is, the number of shrinkage cracks will be reduced as the soil thickness increases. Furthermore, the proposed method could simulate well the curling behaviour observed in experiments. These numerical results suggest that SPH is a promising approach to simulate soil cracking. However, further developments of the soil constitutive model and the SPH algorithm are needed to improve the accuracy of the proposed numerical approach.